# Optical Anisotropy and Pinning of the Linear Polarization of Light in Semiconductor Microcavities

Ł. Kłopotowski[1], M. D. Martín[1], A. Amo[1], L. Viña[1], I.A. Shelykh[2,3], M.M. Glazov[4], G. Malpuech[2], A.V. Kavokin[5], and R. André[6]

[1]SEMICUAM, Departamento de Física de Materiales, Universidad Autónoma de Madrid, Cantoblanco, E28049 Madrid, Spain

[2]LASMEA, CNRS, Université Blaise-Pascal Clermont Ferrand II, 24, av des Landais, 63177, Aubière, France.

[3] St. Petersburg State Polytechnical University, 29, Politechnicheskaya, 195251, St-Petersburg, Russia.

[4] A.F. Ioffe Physico-Technical Institute, RAS, 26, Politechnicheskaya, 194021, St-Petersburg, Russia.

[5] Department of Physics and Astronomy, University of Southampton, SO17 1BJ Southampton, UK

[6] Lab. Spectrométrie Physique (CNRS), Université Joseph Fourier 1, F-38402 Grenoble, France

**Abstract**

We report a strong experimental evidence of the optical anisotropy in a CdTe-based microcavity: the polarization of light is pinned to one of the crystallographic axes independently on the polarization of the excitation. The polarization degree depends strongly on the excitation power, reaching almost 100 % in the stimulated regime. The relaxation time of the polarization is about 1 ns. We argue that all this is an effect of a splitting of the polariton doublet at k=0. We consider different sources for the splitting and conclude that the most likely one is optical birefringence in the mirrors and/or the cavity.

Semiconductor microcavities containing quantum wells (QWs) have attracted the attention of the scientific community since 1990s when the strong light-matter coupling regime has been experimentally documented [1]. In 1996 the concept of polariton laser based on a microcavity has been proposed [2]. In such a laser, light would be emitted spontaneously by a Bose-Einstein condensate (BEC) of exciton-polaritons. In 2000 a clear evidence for stimulated scattering of exciton-polaritons has been obtained [3], which indicated that polariton BEC in microcavities is indeed possible. Very recently, several claims for observation of polariton BEC have been made [4,5], so that the perspective of realization of polariton lasers in the near future looks realistic.

In the quest for polariton lasing, one of very important issues is the polarization dynamics in microcavities. Recently, a spontaneous build-up of the linear polarization in the emission of polariton lasers has been predicted theoretically [6]. According to this theory, the direction of linear polarization would be spontaneously chosen by the system, and it would vary randomly from one experiment to another. In the present work we show experimentally that the linear polarization of emission of polariton lasers can be pinned to one of crystallographic axes. This pinning comes from the optical anisotropy of microcavities which may be caused either by a small birefringence in the mirrors and cavity, by the exciton localization at QW interfaces or by the QW intrinsic anisotropy. A careful analysis allows us to conclude that the first option is the most likely one.

Spin relaxation of exciton-polaritons is correlated with their momentum relaxation in microcavities. This makes microcavities a unique laboratory for studies of spin-polarized bosons. This correlation leads in certain cases to conservation and even build-up of a given polarization in a given quantum state [7]. Recent experiments on polarization dynamics of the photoluminescence (PL) of microcavities excited resonantly or non-resonantly by a polarized laser light have revealed a number of interesting effects including the self-induced Larmor precession [8], quantum beats between dark excitons and polaritons [9], inversion of the polarization degree [10, 11], etc. These effects have been theoretically described in a series of papers [11-13] using various techniques. In most of these papers, the microcavities have been considered as optically isotropic objects having a perfect cylindrical symmetry. Only in a very recent paper [14] it has been suggested that some built-in anisotropy may affect the polarization relaxation of polaritons.

The present work demonstrates that in the case of strong negative detuning between the cavity photon mode and the exciton resonance the optical anisotropy governs the polarization of the emission of the cavity both in the spontaneous and the stimulated regimes. Based on this fact, we expect that the linear polarization of emission of future polariton lasers will not be random but will be pinned to one of the in-plane crystal axes. The life-time of the pinned linear polarization is found to be of the order of a few nanoseconds.

We study experimentally the dynamics of the linear polarization degree of the PL from a semiconductor microcavity after non-resonant, linearly-polarized, pulsed excitation. The sample is a $Cd_{0.4}Mg_{0.6}Te$ $\lambda$-cavity with top (bottom) distributed Bragg reflectors consisting of 17.5 (23) pairs of alternating $\lambda/4$-thick layers of $Cd_{0.4}Mg_{0.6}Te$ and $Cd_{0.75}Mn_{0.25}Te$. In each of the antinodes of the electromagnetic field there are two CdTe QWs of 90 Å thickness. Strong radiation-matter interaction in this structure leads to a Rabi splitting of 10 meV at 5 K. The measurements are performed at a photon-exciton detuning ($\delta$) of -15 meV and a temperature of 5 K. The PL is non-resonantly excited, above the mirror's stop-band (i.e. approximately 62 meV above the cavity resonance), with 2 ps pulses. The excitation arrives to the sample under a small angle (3º) and the PL emitted along the normal to the surface (±1º) is detected using a spectrograph coupled to a streak-camera (energy resolution 0.25 meV, time resolution 10 ps). The excitation is linearly polarized, either horizontally or vertically, and the PL is analysed into its two linearly-polarized components, rendering the linear polarization degree $P_L = (I_\parallel - I_\perp)/(I_\parallel - I_\perp)$, where $I_{\parallel/\perp}$ denotes the intensity of the PL component that is parallel/perpendicular to the excitation. We will concentrate our analysis on the PL arising from the lower polariton branch and will present results obtained under two different excitation conditions, below and above the stimulated scattering threshold (SST, 45 W/cm$^2$), i.e. in the spontaneous and the stimulated scattering regimes, respectively.

Figures 1 (a, b) display the time evolution of the two linearly-polarized components of the PL (semi-logarithmic scale) after horizontally polarized excitation for excitation densities below (6 W/cm$^2$, spontaneous) and above (56 W/cm$^2$, stimulated) the SST, respectively. The decay time amounts to 130 ps/15 ps in the spontaneous/stimulated regime. It can be clearly seen in both figures that the intensity of the vertically polarized emission (dashed line) is larger than that of the horizontally polarized one (solid line) and that this difference of the intensities is visible all through the duration of the emission. This becomes more evident looking at the time evolution of the degree of linear polarization of the emission (obtained from the time evolution traces), which is displayed in figures 1 (c, d), for the same conditions as in figures 1 (a, b). In the spontaneous regime (fig. 1(c)), a build up of a negative polarization is observed during the first 50 ps, after which a net polarization of ~ 35% is recorded as long as there is any measurable signal. In a similar way, a polarization degree of − 65% is obtained for a larger excitation density (fig. 1 (d)) once in the stimulated regime, in which $P_L$ reaches its maximum value (~ 90%). This remarkable enhancement of the linear polarization degree is due to the bosonic stimulation effect, triggered by the large occupation numbers of the ground state obtained in the stimulated scattering regime.

In both regimes, the decay time of the linear polarization degree is of the order of 1 ns, i.e. much longer than the intensity decay time. The inset shows the time evolution of the circular polarization degree of the PL obtained under similar conditions as in the experiment described above but using circularly polarized excitation/detection: it shows a fast decay on a time scale of 40 ps and then becomes negligible. Thus, surprisingly, the linear polarization decay time is much longer than all the other characteristic times of the system. Furthermore, the fact that $P_L$ is negative implies a 90º rotation of the polarization plane of the emission with respect to that of the excitation.

To find out more about this 90º rotation of the polarization we rotated the excitation plane by 90º and excited the PL with vertically polarized light. Figure 2 summarizes the results obtained in the stimulated regime (85 W/cm$^2$). Figure 2 (a) displays the time evolution of the two linearly polarized components of the PL after <u>horizontally</u> polarized excitation, showing a behaviour completely analogous to the one described above: the vertically polarized emission (dashed line) is stronger than the horizontally polarized one (solid line). In a similar fashion, figure 2 (b) shows the time evolution traces obtained under similar excitation conditions but with the polarization of the <u>excitation rotated by 90º</u>. Yet, the vertically polarized component of the PL (dashed line) is stronger all through the duration of the emission. These experimental results show that the polarization of the emission is pinned to one of the crystallographic axes of the structure. This pinning has been confirmed by supplementary experiments performed with the sample rotated 90º, returning to same spot and using similar excitation conditions. In this case, we have found also that the preferential orientation for the polarization of the emission is rotated by 90º, thus confirming the pinning along the same crystal axis as before.

We interpret these striking effects, i.e. the long decay time of $P_L$ and the pinning of the linear polarization of the PL, as a consequence of a splitting of the polariton ground state. This splitting can arise from either a splitting of the exciton resonance into a linearly polarized radiative doublet or from a splitting of the photon eigen-modes of the cavity polarized vertically and horizontally. It is worth mentioning that the longitudinal-transverse splitting of exciton-polaritons, which is responsible for polariton spin relaxation in the excited states [12] is always zero at k = 0.

Before discussing the possible origins of the splitting, $\Delta\varepsilon$, of the two linearly polarized polariton states at k=0, let us see how it may affect the polarization properties of the light emission of the system. To get a simple idea of the scale of the effect, let us consider a polariton gas in thermal equilibrium. Of course, it is never the case for exciton polaritons, for which the finite lifetime and the bottleneck effect are of crucial importance, but a simple thermodynamic treatment will allow us to understand qualitatively the observed phenomena. In the Boltzmann limit, the linear polarization degree of the emission from the ground state $P_L$ can be estimated as

$$P_L = \tanh\left(\frac{\Delta\varepsilon}{2k_B T}\right). \qquad (1)$$

Taking $\Delta\varepsilon = 200$ µeV, as obtained from our data, and a temperature of 5K we obtain $P_L \approx 0.46$, which is close to what we observe in the linear regime. The splitting, $\Delta\varepsilon$, can be hinted in our time-resolved measurements. The inset in Figure 2 displays the two linearly polarized components of the time-integrated PL for the weakest excitation used in our experiments. One can see a small splitting between the two spectra, which cannot be measured with high accuracy because it is below the energy resolution (250 µeV) and it is much smaller than the linewidth of the PL. Additional experiments performed under continuous-wave excitation confirm the existence of the energy splitting between the two linearly polarized components of the PL and show that the component lying at higher energies has always a larger intensity than that of the counter-polarized one, indicating the non-thermal origin of the polarization. These experiments also show that $\Delta\varepsilon$ is very robust. The linear polarization and the splitting have a similar temperature dependence, monotonically decreasing with T and both vanishing at T ~ 100 K, thus confirming the direct relationship between the energy splitting and the observed linear polarization degree.

Above the stimulation threshold, due to the bosonic nature of exciton-polaritons, the degree of the linear polarization is significantly enhanced compared to what Eq. (1) yields. If the number of the particles in the ground state is $N$, we obtain

$$P_L = \frac{1}{N}\left[\frac{1}{A-1} - \frac{1}{Ae^{\Delta\varepsilon/kT}-1}\right], \quad A = \frac{(N+1)\left(e^{-\Delta\varepsilon/kT}+1\right) + \sqrt{N(N+2)\left(e^{-\Delta\varepsilon/kT}-1\right)^2 + \left(e^{-\Delta\varepsilon/kT}+1\right)^2}}{2N} \quad (2)$$

In the limit $N \gg 1$ these formulae can be sufficiently simplified and the linear polarization degree reads

$$P_L = 1 - \frac{4}{N\left(e^{\Delta\varepsilon/kT}-1\right)}. \quad (3)$$

It is seen that even for a very small splitting the degree of polarization can be almost unity if the number of the particles in the ground state is large enough. This is in agreement with our experimental observations showing a drastic increase of $P_L$ above the stimulation threshold when a large occupation of the ground state is obtained and the bosonic effects play a major role, although the intensities of the emission are not determined by thermal occupation of the split levels.

Now let us discuss the possible origin of the ground state splitting. The polariton eigen state is a combination of an excitonic and photonic eigen state and the intermixing depends on the exciton-photon coupling constant and the detuning between bare exciton and photon modes. In principle, the splitting could arise from a splitting of the exciton state and a different exciton-photon coupling constant in the two linear polarizations. The splitting of the exciton ground state is forbidden by symmetry reasons in an ideal symmetric QW, but becomes possible if the QW is asymmetric or if its interfaces have fluctuation islands oriented along the crystal axes [15]. The latter mechanism can hardly be the dominant one: we observe a drastic increase of $P_L$ above the stimulation threshold, which implies large occupation numbers that are incompatible with localized states.

Let us now discuss the remaining possibility of the free exciton energy splitting, linked with an asymmetry along the growth axis (z) of the QWs embedded in the cavity [16]. Under these circumstances, the exciton state becomes preferentially localized at one of interfaces which has a lower symmetry than the bulk crystal. This may happen due to built-in electric fields [17] or simply due to the different scale of interface roughness at two interfaces. Such localization would break the $D_{2d}$ symmetry of the exciton state in a QW and reduce it to the $C_{2v}$ symmetry. In this case $x$ // [110] and $y$ // [1-10] axes are no longer equivalent [18, 19], and the exciton states split into $x$- and $y$-polarized states. Microscopically, an electric field (either external or built-in) induces a mixing of the heavy hole and light hole wave-functions ($\psi_{\pm 3/2}^{hh}(z)$ and $\psi_{\mp 1/2}^{lh}(z)$ respectively) at the interfaces, which lifts the degeneracy of the exciton ground state.

The value of splitting between $x$- and $y$-polarized states has been evaluated in [18]:

$$\Delta \varepsilon_{ex} = \varepsilon_0 \frac{16 a_0^3}{\sqrt{3} \pi a_B^2} \int \psi_{\pm 3/2}^{hh}(z) \psi_{\mp 1/2}^{lh}(z) |\psi^e(z)|^2 \, dz, \qquad (4)$$

where $\varepsilon_0$ is the short-range exchange interaction constant, $a_0$ is the lattice constant, $a_B$ is the 2D exciton Bohr radius and $\psi^e(z)$ is the electron envelope function. This splitting can achieve 50-80 $\mu eV$ for realistic QW parameters.

Besides, the low symmetry of the structure induces a variation of the exciton oscillator strength in $x$- and $y$- linear polarizations. This is because the interband matrix element of the optical transition for zinc-blend lattice semiconductors has a contribution linear in $k_z$ [17]

$$\left\langle c, \pm \frac{1}{2} \middle| \mathbf{ep} \middle| v, \pm \frac{3}{2} \right\rangle = P e_\pm - Q k_z e_\mp, \qquad (5)$$

where $P$ is the momentum matrix element calculated for $k_z = 0$, $Q$ is a constant describing **k**-linear contribution, and **e** is the photon polarization vector ($e_\pm = e_x \pm i e_y$). Eq. (5) is written in the conduction band representation. According to Eq. (5) the matrix elements of exciton-light coupling in $x$ and $y$ polarizations are different. This results in the difference of the Rabi splittings for different linear polarizations

$$\Delta V_R = V_R \wp , \tag{6}$$

where $V_R$ is the Rabi splitting for a symmetric QW and $\wp$ the linear polarization degree of the PL from a single QW in the spontaneous regime, can be estimated as:

$$\wp \approx \frac{2I_2}{\sqrt{3}I_1} - \frac{2QI_3}{a_0 P I_1}, \tag{7}$$

where $I_1 = \int \psi^e(z) \psi^{hh}_{3/2}(z) dz$, $I_2 = \int \psi^e(z) \psi^{lh}_{-1/2}(z) dz$, and $I_3 = a_0 \int \psi^e(z) \frac{d\psi^{hh}_{3/2}(z)}{dz} dz$. Two contributions into Eq. (7) come from the light-heavy hole mixing at interfaces and from the linear in $k_z$ term in Eq. (5), respectively. Together with the exciton splitting given by Eq. (4) this contributes to the splitting between the linear polarized states at the lower polariton branch, which reads:

$$\Delta \varepsilon_{pol} = \frac{\Delta \varepsilon_{ex}}{2} \left( 1 + \frac{\delta}{\sqrt{\delta^2 + V_R^2}} \right) + \frac{\wp V_R^2}{2\sqrt{\delta^2 + V_R^2}} \tag{8}$$

where δ is a detuning parameter that is taken to be positive if the lowest excitonic band lies below the photonic band, and negative in the opposite case. One can see from Eq. (8) that at large negative detuning, the polariton splitting is strongly reduced with respect to the pure exciton splitting. In our case the polariton splitting will not be more than 10-20 $\mu eV$. This is not enough to describe the linear polarization degree of emission that we observe in the spontaneous regime.

After ruling out the previously discussed source for the splitting, one is left with the effect of the photon eigen states. Since the data presented here have been obtained under a large negative detuning (δ = -15 meV), the polariton ground state is about 75% photon-like and therefore any effects associated with the photonic part of the polariton can be enhanced. Let us consider a small splitting between the bare photon modes of our microcavity in horizontal and vertical polarizations. This would mean that the Bragg mirrors or the cavity itself are slightly birefringent. The frequency $\omega_c$ of the uncoupled cavity mode is inversely proportional to the refractive index of the cavity material $n_c$. In an ideal $\lambda$-microcavity $\omega_c = 2\pi c / n_c L_c$, where $L_c$ is the cavity width. Thus, a small change of $n_c$ leads to a variation of the cavity frequency given by

$$\Delta \omega_c \approx -\frac{\Delta n_c}{n_c} \omega_c. \qquad (9)$$

To obtain a polariton splitting of 200 μeV, the refractive index of the cavity should vary by about 0.02% between horizontal and vertical polarizations. Such a small variation can be a result of a weak uniaxial strain in-plane of the cavity. We believe it is the most likely possibility.

In conclusion, we have demonstrated experimentally a pinning of the linear polarization degree in non-resonantly pumped microcavities. The polarization degree of the emission is as high as 35% in the spontaneous regime and increases up to almost 100% above the stimulation threshold due to the polariton condensation. The pinning of the polarization is caused by a splitting of the polariton ground state, which is due to a slight birefringence of the cavity or the mirrors as follows from our theoretical analysis. We expect that in future polariton lasers the linear polarization of emission will be stabilized and linked to one of the crystal axes due to this pinning effect.

We thank F. Laussy and K.V. Kavokin for useful discussions. The work was partially supported by the Spanish MEC (MAT2005-01388 and NAN2004-09109-C04-04), the CAM (S-0505/ESP-0200), the RFBR and the "Marie-Curie" MRTN-CT-2003-503677.

*List of references*

1. C. Weisbuch, M. Nishioka, A. Ishikawa, and Y. Arakawa, Phys. Rev. Lett. **69**, 3314 (1992).

2. A. Imamoglu, J. R. Ram, Phys. Lett. A **214**, 193, (1996).

3. P.G. Savvidis, J.J. Baumberg, R. M. Stevenson, M.S. Skolnick, D.M. Whittaker and J.S. Roberts, Phys. Rev. Lett. **84**, 1547 (2000).

4. H. Deng, G. Weihs, C. Santori, J. Bloch and Y. Yamamoto, Science **298**, 199 (2002).

5. M. Richard, J. Kasprzak, R. André, R. Romestain, Le Si Dang, G. Malpuech, and A. Kavokin, Phys. Rev. **B 72**, 201301 (2005).

6. F. P. Laussy, I. A. Shelykh, G. Malpuech, and A. Kavokin, Phys. Rev. **B 73**, 035315 (2006).

7. A. I. Tartakovskii, V. D. Kulakovskii, D. N. Krizhanovskii, M. S. Skolnick, V. N. Astratov, A. Armitage, and J. S. Roberts, Phys. Rev. **B 60**, R11293 (1999). M.D. Martin, G. Aichmayr, L. Viña, R. André, Phys. Rev. Lett. **89**, 77402 (2002).

8. P. G. Lagoudakis, P. G. Savvidis, J. J. Baumberg, D. M. Whittaker, P. R. Eastham, M. S. Skolnick, and J. S. Roberts, Phys. Rev. **B 65**, 161310 (2002).

9. I.A. Shelykh, L. Vina, A.V. Kavokin, N.G. Galkin, G. Malpuech, R. Andre, Solid State Comm. **135**, 1 (2005).

10. M. D. Martín, L. Viña, J. K. Son, and E. E. Mendez, Solid State Commun. **117**, 267 (2001); P. Renucci, T. Amand, X. Marie, P. Senellart, J. Bloch, B. Sermage, and K. V. Kavokin, Phys. Rev. **B 72**, 075317 (2005).

11. A.V. Kavokin, P.G. Lagoudakis, G. Malpuech, J.J. Baumberg, Phys. Rev. **B 67**, 195321 (2003).

12. K.V. Kavokin, I. Shelykh, A.V. Kavokin, G. Malpuech, and P. Bigenwald, Phys. Rev. Lett., **92**, 017401 (2004).

13. I. Shelykh, G. Malpuech, K. V. Kavokin, A. V. Kavokin, and P. Bigenwald, Phys. Rev. **B 70**, 115301 (2004).

[14] D. N. Krizhanovskii, D. Sanvitto, I. A. Shelykh, M. M. Glazov, G. Malpuech, D. D. Solnyshkov, A. Kavokin, S. Ceccarelli, M. S. Skolnick, and J. S. Roberts, Phys. Rev. **B 73**, 073303 (2006).

[15] S.V. Goupalov, E.L. Ivchenko, and A.V. Kavokin. *JETP* **86**, 388 (1998).

[16] A. Kudelski, A. Golnik, J. A. Gaj, F. V. Kyrychenko, G. Karczewski, T. Wojtowicz, Yu. G Semenov, O. Krebs and P. Voisin, Phys. Rev. **B 64**, 045312 (2001).

[17] See e.g. Chapter 3 of E.L. Ivchenko, *Optical Spectroscopy of semiconductor nanostructures*, Alpha Science, Harrow UK (2005).

[18] I.L. Aleiner, E.L. Ivchenko, *Pis'ma Zh. Exper. Teor. Fiz.* **55**, 662 (1992) [*JETP Letters* **55**, 692 (1992).], E.L. Ivchenko, A. Yu. Kaminski and U. Roessler, *Phys. Rev.* **B 54**, 5852 (1996).

[19] E.L. Ivchenko, A.A. Toropov, P. Voisin, *Fiz. Tverd. Tela* **40**, 1925 (1998) [*Phys. Solid State* **40**, 1748 (1998)], A.A. Toropov, *et al.*, *Phys. Rev.* **B 63**, 035302 (2000).

13
*List of figure captions*

**Figure 1**: (a) and (b): Time-evolution of the PL, in semi-logarithmic scale, of the lower polariton branch in the spontaneous (6 W/cm$^2$) and stimulated (56 W/cm$^2$) regimes, respectively. Solid/dashed lines are linearly H and V -polarized signals. (c) and (d): temporal profiles of the linear polarization degree, $P_L$, in the spontaneous and stimulated regimes, respectively. The inset shows the decay of the circular polarization degree under circular-polarized pumping in the spontaneous regime.

**Figure 2**: Time evolution traces of the PL, in a semi-logarithmic scale, of the lower polariton branch in the stimulated regime (85 W/cm$^2$) after linearly polarized excitation along the (a) horizontal (H) and (b) vertical (V) direction. Solid/dashed lines are linearly H and V -polarized signals. The inset displays the two linearly polarized components (horizontal/vertical in black/light-grey shade) of the time-integrated PL in the spontaneous regime (3 W/cm$^2$). The dashed white lines are guides to the eye.

Figure 1

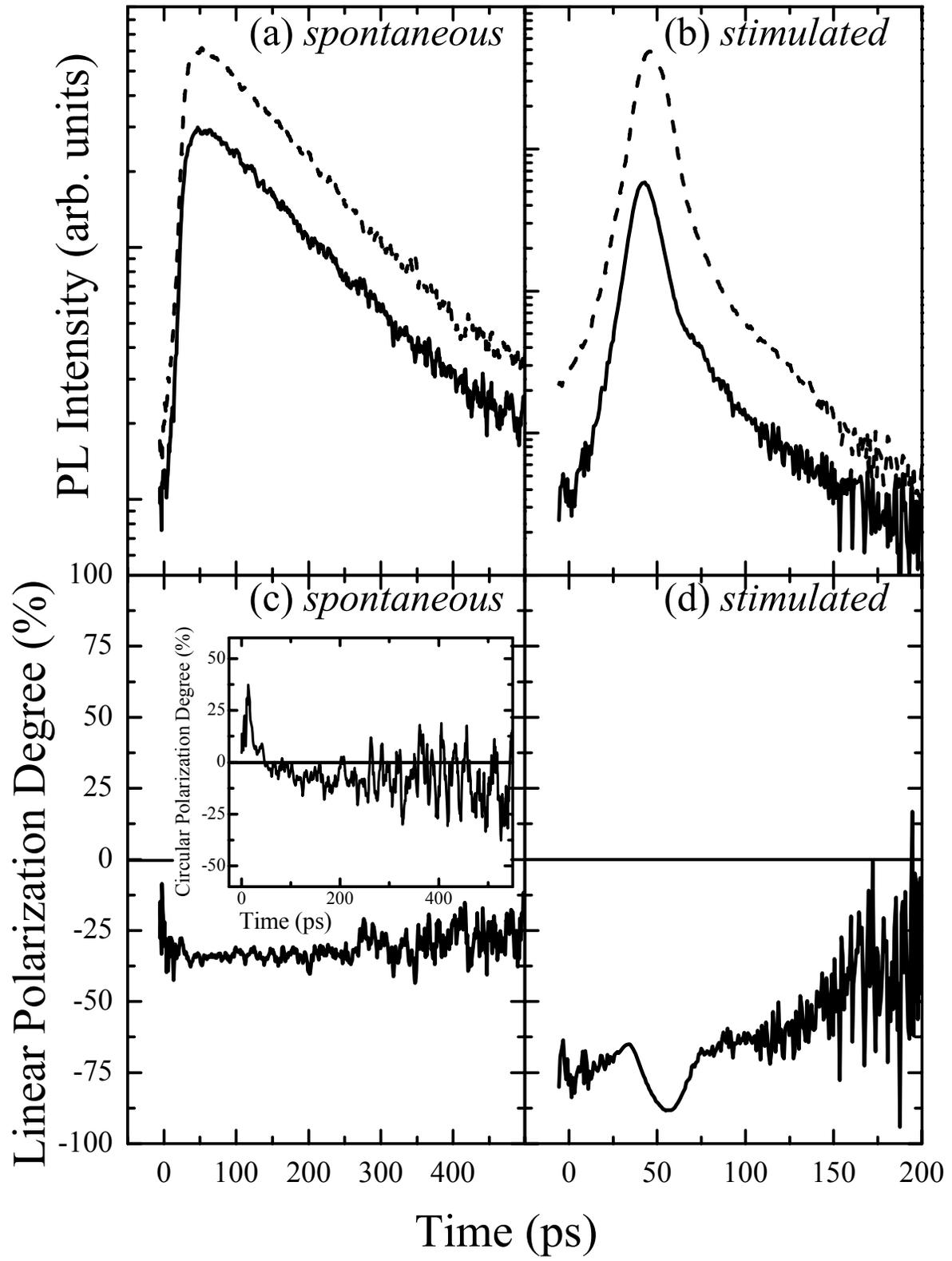

Figure 2

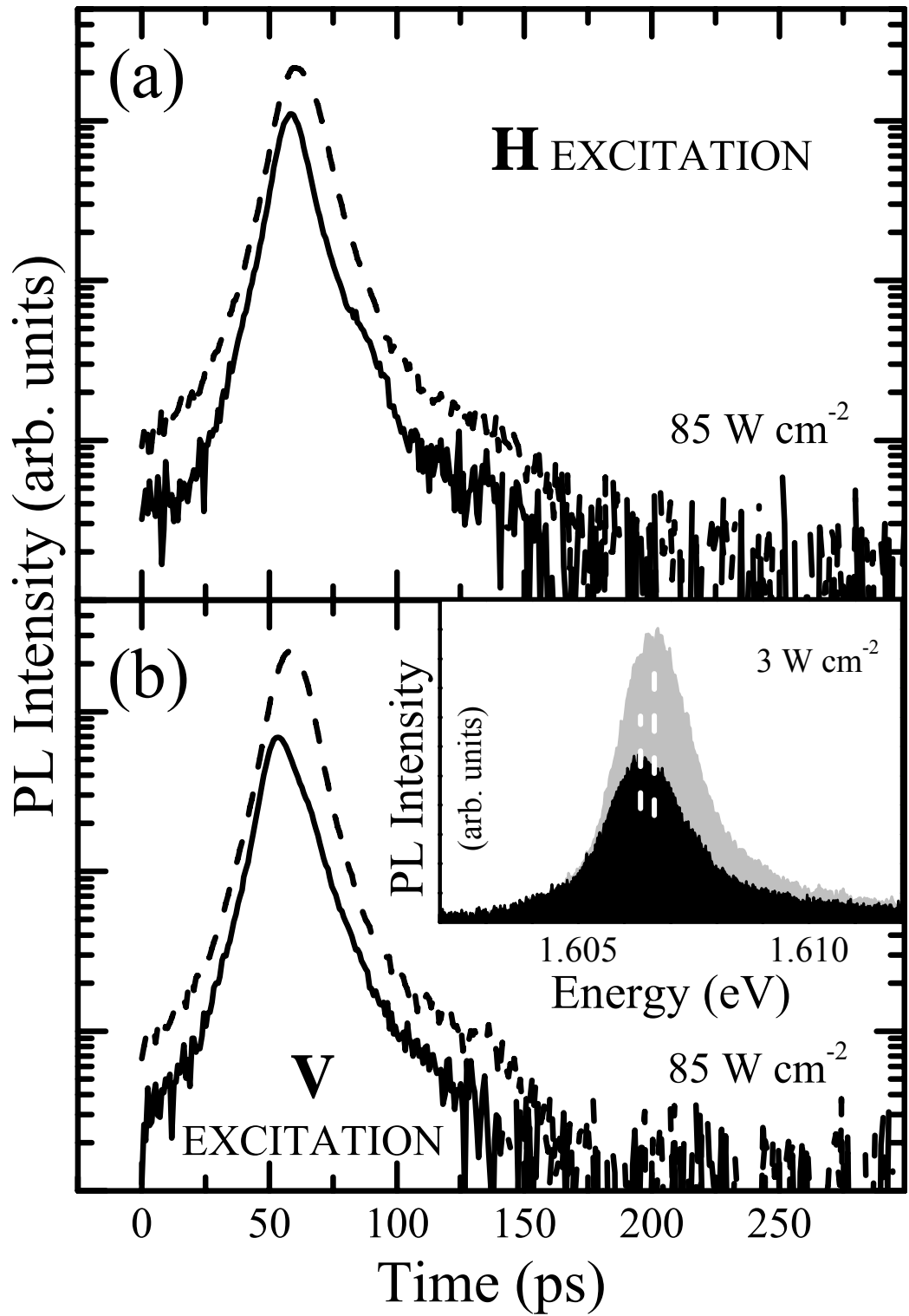